\documentclass[twoside,english,british]{elsarticle}
\usepackage[T1]{fontenc}
\usepackage[latin9]{inputenc}
\usepackage{color}
\usepackage{amstext}
\usepackage{amssymb}
\usepackage{graphicx}
\usepackage{esint}

\makeatletter
\journal{Example: Nuclear Physics B}

\makeatother

\usepackage{babel}
\begin{document}

\title{Effects of nitrogen-doping configurations with vacancies on conductivity
in graphene}

\author[rvt]{T.M.~Radchenko\corref{cor1}}

\ead{tarad@imp.kiev.ua}

\author[rvt]{V.A.~Tatarenko}

\author[focal]{I.Yu.~Sagalianov}

\author[focal]{Yu.I.~Prylutskyy}

\cortext[cor1]{Corresponding author}

\address[rvt]{Department of Solid State Theory, G.V. Kurdyumov Institute for Metal
Physics of NASU, 36 Acad. Vernadsky Blvd., Kyiv, Ukraine}

\address[focal]{Taras Shevchenko National University of Kyiv, 64 Volodymyrska Str.,
Kyiv, Ukraine}
\begin{abstract}
We investigate electronic transport in the nitrogen-doped graphene
containing different configurations of point defects: singly or doubly
substituting N atoms and nitrogen--vacancy complexes. The results
are numerically obtained using the quantum-mechanical Kubo--Greenwood
formalism. Nitrogen substitutions in graphene lattice are modelled
by the scattering potential adopted from the independent self-consistent
\textit{ab initio} calculations. Variety of quantitative and qualitative
changes in the conductivity behaviour are revealed for both graphite-
and pyridine-type N defects in graphene. For the most common graphite-like
configurations in the N-doped graphene, we also consider cases of
correlation and ordering of substitutional N atoms. The conductivity
is found to be enhanced up to several times for correlated N dopants
and tens times for ordered ones as compared to the cases of their
random distributions. The presence of vacancies in the complex defects
as well as ordering of N dopants suppresses the electron--hole asymmetry
of the conductivity in graphene.\end{abstract}
\begin{keyword}
graphene \sep quantum transport \sep point defects \PACS 81.05.ue
\sep 72.80.Vp \sep 72.10.Fk \MSC[2014]81U35
\end{keyword}
\maketitle

\section{Introduction}

Among the currently known and already experimentally implemented or
prospective substitutional dopants in graphene films (\textit{e.g.},
B \citep{Sheng2012,Norimatsu2012,Wang2013,Gebhardt2013,Panchakarla2009,Zheng2010,Zhao2013},
N \citep{Panchakarla2009,Zheng2010,Zhao2013,Guo2010,Wei2009,Joucken2012,Lv2012,Fujimoto2011,Hou2012,Zhao2011},
Al--S \citep{Dai2009,Dai2010,Zhang2009}, Sc--Zn \citep{Santos2010_,Santos2010,Santos2008,KrasheninnikovPRL},
Pt \citep{KrasheninnikovPRL,Gan}, Au \citep{KrasheninnikovPRL,Gan},
Bi \citep{Akt=0000FCrk}), nitrogen (along with boron) is an archetypical
natural candidate because its incorporation in graphene lattice requires
minor structural perturbations due to its atomic size close to C.
The N doping offers an effective way to tailor the properties of graphene
and thereby makes it a promising material for applications in field-effect
transistor devices, solar and fuel cells, lithium ion batteries, ultracapacitors,
biosensing, field emission, transparent electrodes, or high-performance
photocatalysts (see Ref. \citep{Wang2012} and references therein).

\begin{figure*}
\includegraphics[width=1\textwidth]{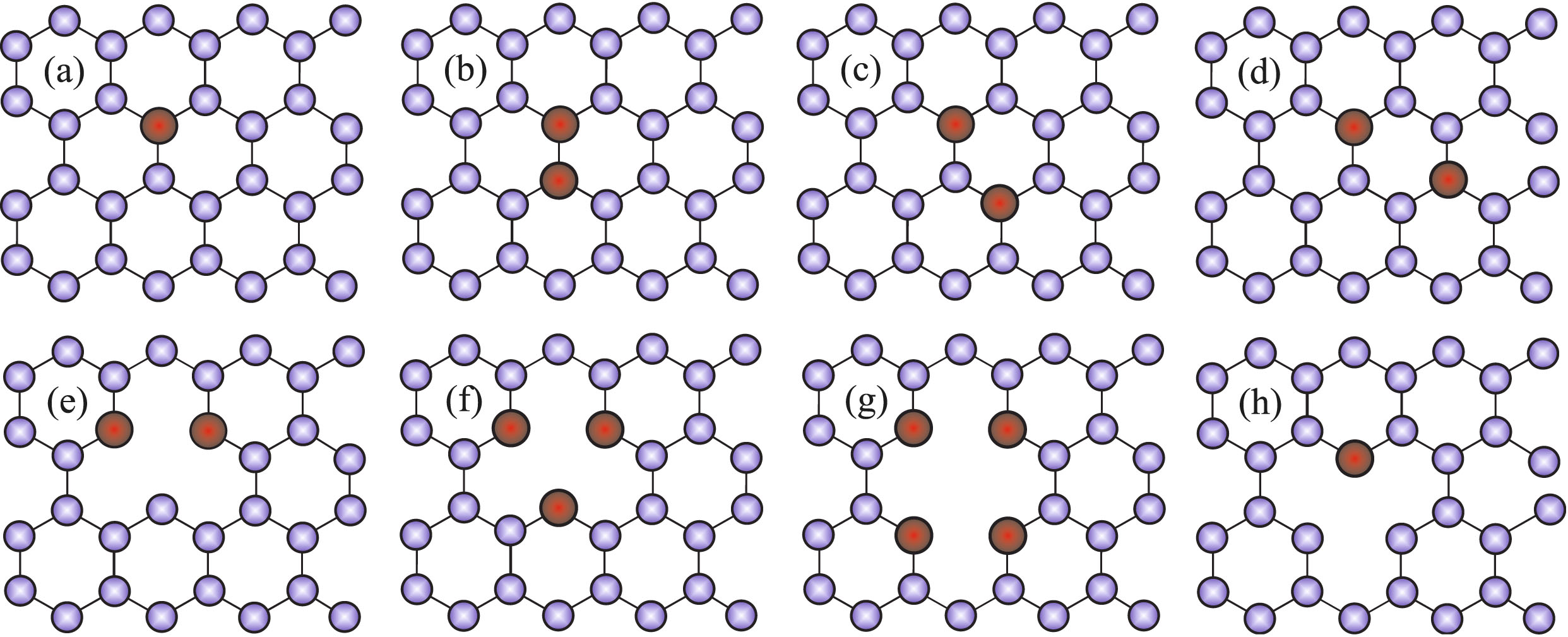}

\caption{(Colour online) (a)--(d) Graphite- (substitutional) and (e)--(h) pyridine-type
defects in graphene lattice. N atoms substitute C ones (a) singly,
(b)--(d) doubly or form complex (e) dimerized, (f) trimerized, (g)
tetramerized, or (h) monomeric defects with vacancies. Doubly substituting
N atoms belong to the same sublattice (c) or different ones (b), (d).}

\label{Fig_Configurations}
\end{figure*}

A series of experiments on structural variety of N-related defects
in graphene \citep{Zheng2010,Zhao2013,Guo2010,Wei2009,Joucken2012,Lv2012}
showed that its rich electronic properties are dependent on how the
N-doping configurations are formed: as single (Fig. \ref{Fig_Configurations}(a))
or double (Figs. \ref{Fig_Configurations}(b)--\ref{Fig_Configurations}(d))
substitutions or as nitrogen--vacancy complexes comprising monovacancies
(Figs. \ref{Fig_Configurations}(e), \ref{Fig_Configurations}(f),
\ref{Fig_Configurations}(h)) or divacancies (Fig. \ref{Fig_Configurations}(g)).
Single and double substitutions give rise to the electron-donor-like
states and then to $n$-type doping, while complex defects with vacancies
exhibit a hole-acceptor-like character, \textit{i.e.} $p$-type doping
\citep{Zheng2010,Wei2009,Joucken2012}. Both graphite- and pyridine-type
defects in Fig. \ref{Fig_Configurations} can be observed in graphene
films, which are fabricated by chemical vapour deposition growth on
different substrates \citep{Zhao2013,Wei2009,Lv2012}. Formation-energy
calculations \citep{Fujimoto2011} indicate that graphitic defects
in Fig. \ref{Fig_Configurations}(a) are energetically favoured (stable)
among the possible N-doping configurations in Fig. \ref{Fig_Configurations}.
Pyridine-like configurations in Figs. \ref{Fig_Configurations}(e)--\ref{Fig_Configurations}(h)
have higher formation energy, but are stable in the presence of both
doping N atoms and vacancies attracting each other and increasing
probability of their mutual generation \citep{Hou2012}. The first
numerical study of charge transport in N-doped graphene \citep{Lherbier2008}
deals with the most simple case of random distribution of singly substituting
N atoms (Fig. \ref{Fig_Configurations}(a)). However, in order to
regulate the transport properties of graphene by chemical N-doping,
it is important to consider all configurations of the N-related defects
currently revealed in experiments.

In a given paper, we report on how such diverse N-doping configurations
affect the conductivity in single-layer graphene sheet, using an exact
numerical technique based on the Kubo--Greenwood formalism appropriate
for realistic samples with millions of atoms (the size of our computational
domain is 1700000 sites that corresponds to $210\times210$ nm$^{2}$).
We also focus on random, correlated, and ordered distributions of
N dopants in one of the most common doping configurations, which is
found \citep{Joucken2012} to be singly substituting N atom in Fig.
\ref{Fig_Configurations}(a).

\section{Tight-binding model along with Kubo--Greenwood formalism}

To investigate charge transport in the N-doped graphene, a real-space
numerical implementation within the Kubo--Greenwood formalism \citep{Roche_SSC,Radchenko 1},
which captures all (ballistic, diffusive, and localization) transport
regimes, is employed. Within this approach, the energy ($E$) and
time ($t$) dependent transport coefficient, $D(E,t)$ \citep{D_t},
is governed by the wave-packet propagation \citep{Roche_SSC,Radchenko 1}:
$D(E,t)=\bigl\langle\Delta\hat{X}^{2}(E,t)\bigr\rangle/t$, where
the mean quadratic spreading of the wave packet along the direction
$x$ reads as \citep{Roche_SSC,Radchenko 1}

\begin{equation}
\bigl\langle\Delta\hat{X}^{2}(E,t)\bigr\rangle=\frac{\text{Tr}[\hat{(X}(t)-\hat{X}(0))^{2}\delta(E-\hat{H})]}{\text{Tr}[\delta(E-\hat{H})]}\label{Eq_Spreading}
\end{equation}
with $\hat{X}(t)=\hat{U}^{\dagger}(t)\hat{X}\hat{U}(t)$---the position
operator in the Heisenberg representation, $\hat{U}(t)=e^{-i\hat{H}t/\hbar}$---the
time-evolution operator, and a standard $p$-orbital nearest-neighbour
tight-binding Hamiltonian $\hat{H}$ is \citep{PeresReview,DasSarmaReview}

\begin{equation}
\hat{H}=-u{\textstyle \sum_{i,i^{\prime}}}c_{i}^{\dagger}c_{i^{\prime}}+{\textstyle \sum_{i}}V_{i}c_{i}^{\dagger}c_{i},\label{Eq_Hamiltonian}
\end{equation}
where $c_{i}^{\dagger}$ ($c_{i}$) is a standard creation (annihilation)
operator acting on a quasiparticle at the site $i$. The summation
over $i$ runs the entire honeycomb lattice, while $i^{\prime}$ is
restricted to the sites next to $i$; $u=2.7$ eV is the hopping integral
for the neighbouring C atoms occupying $i$ and $i^{\prime}$ sites
at a distance $a=0.142$ nm between them; and $V_{i}$ is the on-site
potential describing scattering by the N dopants.

\begin{figure}
\includegraphics[width=0.6\textwidth]{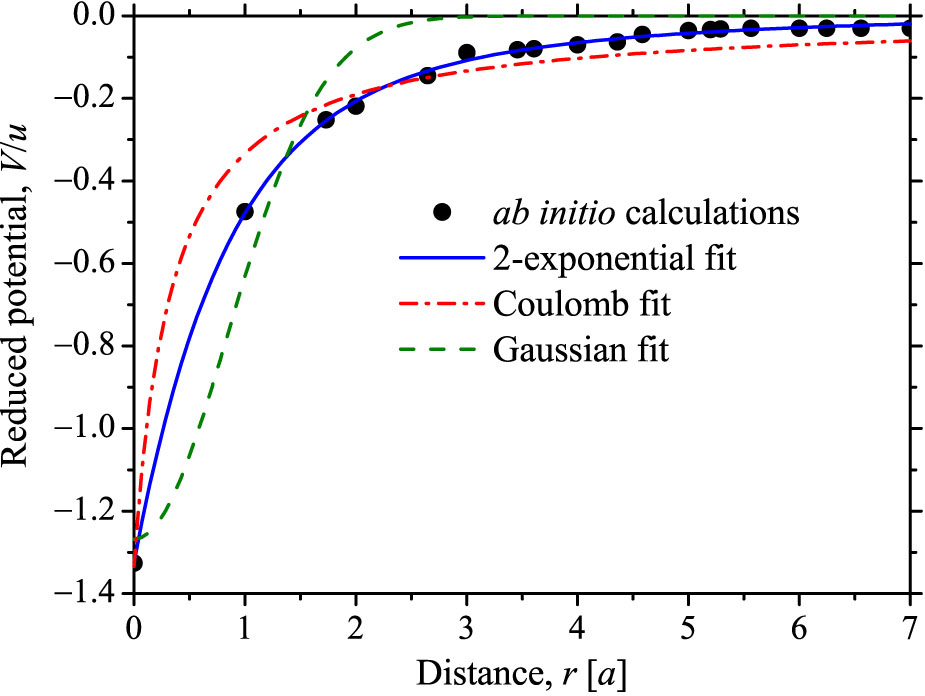}

\caption{(Colour online) Scattering potential ($\bullet$) adopted from Ref.
\citep{Adessi2006} and fitted by different functions:\textcolor{black}{{}
Gaussian-like ($V=Ue^{-\mathbf{\mathrm{\mathit{r}}}^{2}/2\xi^{2}}$
with fitting parameters $U=-1.27u$ and $\xi=0.85a$ standing here
as a maximal potential height and an effective potential radius, respectively),
Coulomb-like ($V=U/(\xi+r)$ with $U=-0.44ua$ and $\xi=0.33a$),
and two-exponential ($V=U_{1}e^{-r/\xi_{1}}+U_{2}e^{-r/\xi_{2}}$
with $U_{1}=-1.07u$, $\xi_{1}=0.79a$, $U_{2}=-0.25u$, $\xi_{2}=2.72a$). }}

\label{Fig_potential}
\end{figure}

The impurity scattering potential in the Hamiltonian matrix is introduced
as on-site energies $V_{i}$ varying with distance $r$ to the impurity
N atom at the site $i$ according to the potential profile $V=V(r)<0$
in Fig. \ref{Fig_potential} adopted from the self-consistent \textit{ab
initio} calculations \citep{Adessi2006}. As fitting shows, this potential
is far from the Coulomb- or Gaussian-like shapes commonly used in
the literature for charged impurities in graphene, \textcolor{black}{while
two-exponential fitting exactly reproduces the potential. Such a scattering
potential presents both short-range and some long-range features }\foreignlanguage{english}{\textcolor{black}{\citep{Lherbier2008}}}\textcolor{black}{.}
A vacancy can be regarded as a site with hopping parameters to other
sites being zero (note that another way to model vacancy at the site
$i$ is $V_{i}\rightarrow\infty$) \citep{Yuan2010}. In our numerical
simulations, we implement a vacancy removing the atom at the vacancy
site. 

The dc conductivity $\sigma$ can be extracted from the diffusivity
$D(E,t)$, when it saturates reaching the maximum value, $\lim_{t\rightarrow\infty}D(E,t)=D_{\max}(E)$,
and the diffusive transport regime occurs. Then the semiclassical
conductivity at a zero temperature is defined as \citep{Roche_SSC,Radchenko 1} 

\begin{equation}
\sigma=e^{2}\tilde{\rho}(E)D_{\max}(E),\label{Eq_sigmaMax}
\end{equation}
where $-e<0$ denotes the electron charge and $\tilde{\rho}(E)=\rho/\Omega=\textrm{Tr}[\delta(E-\hat{H})]/\Omega$
is the density of sates (DOS) per unit area $\Omega$ (and per spin).
The DOS is also used to calculate the electron density as $n_{e}(E)=\int_{-\infty}^{E}\tilde{\rho}(E)dE-n_{\text{ions}},$
where $n_{\text{ions}}=3.9\cdot10^{15}$ cm$^{-2}$ is the density
of the positive ions in the graphene lattice compensating the negative
charge of the $p$-electrons (at the neutrality (Dirac) point of pristine
graphene, $n_{e}(E)=0$). Combining the calculated $n_{e}(E)$ with
$\sigma(E)$, we compute the density dependence of the conductivity
$\sigma=\sigma(n_{e})$.

Note that we do not go into details of numerical calculations of DOS,
$D(E,t)$, and $\sigma$ since details of the computational method
we utilize here (Chebyshev method for solution of the time-dependent
Schrödinger equation, calculation of the first diagonal element of
the Green's function using continued fraction technique and tridiagnalization
procedure of the Hamiltonian matrix, averaging over the N and vacancy
realizations, sizes of initial wave packet and computational domain,
boundary conditions, etc.) are given in Ref. \citep{Radchenko 1}.

\section{Results and discussion}

\begin{figure*}
\includegraphics[width=1\textwidth]{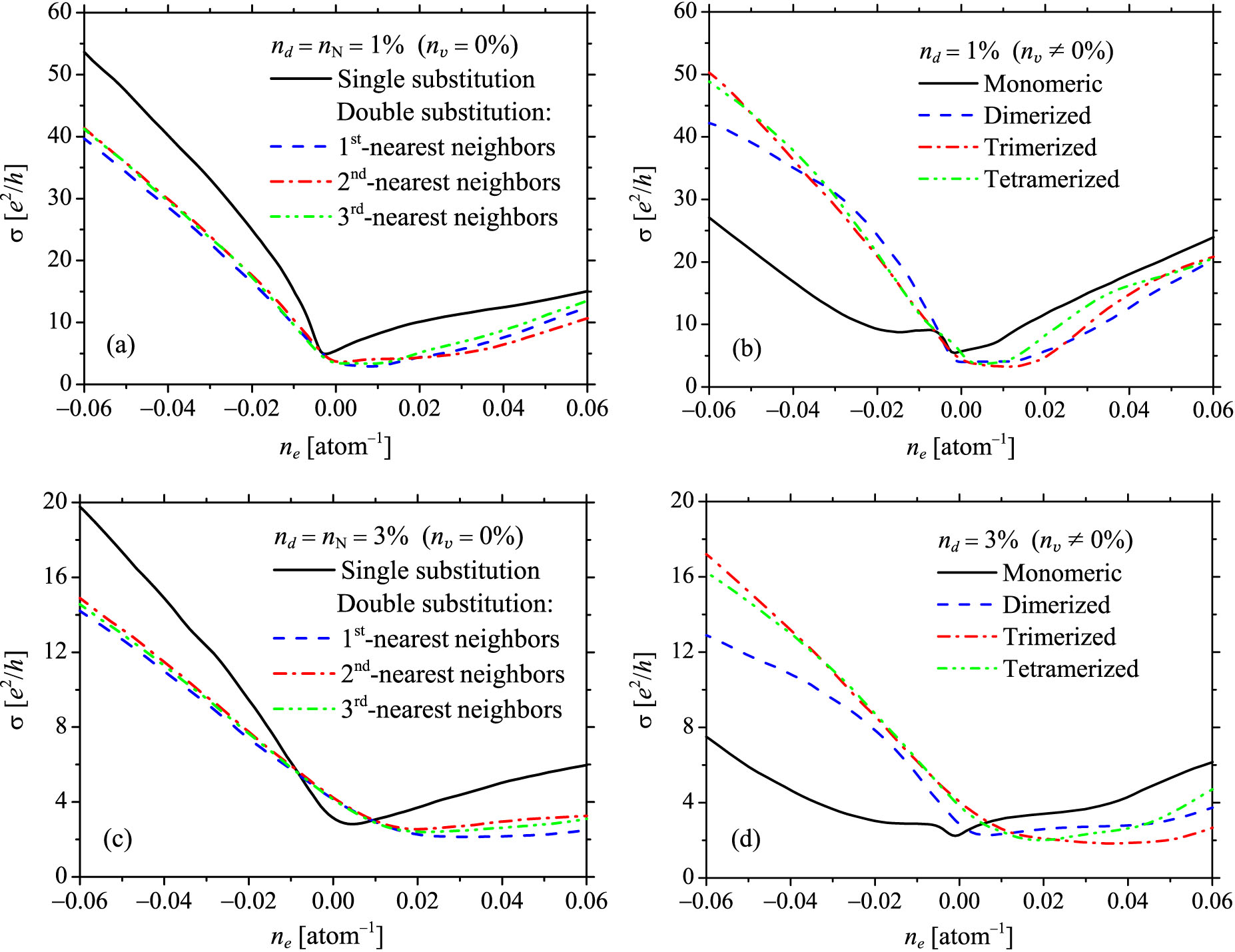}

\caption{(Colour online) Conductivity vs. the electron density for 1\% (a),
(b) and $3\%$ (c), (d) of point N-related defects with graphite-
(a), (c) and pyridine-like (b), (d) arrangements represented in Fig.
\ref{Fig_Configurations}. }

\label{Fig_conductivity}
\end{figure*}

Figure \ref{Fig_conductivity} demonstrates the electron-density dependent
conductivity for defect configurations depicted in Fig. \ref{Fig_Configurations}.
To compare conductivity curves, we chose two representative concentrations
of the point nitrogen-related defects: $n_{d}=n_{\mathrm{N}}+n_{v}=1\%$
and $n_{d}=3\%$, where $n_{\mathrm{N}}$ and $n_{v}$ are nitrogen
and vacancy relative concentrations, respectively. In case of the
graphitic defects, concentration of point defects $n_{d}$ equals
with concentration of substituting N atoms $n_{\mathrm{N}}$ ($n_{d}=n_{\mathrm{N}}$),
since $n_{v}=0$. For the pyridinic defects, the vacancy concentration
$n_{v}\neq0$ that leads to $n_{\mathrm{N}}=n_{v}=n_{d}/2$ for monomeric
defects; $n_{\mathrm{N}}=2n_{d}/3$ and $n_{v}=n_{d}/3$ for dimerized
as well as tetramerized defects; $n_{\mathrm{N}}=3n_{d}/4$ and $n_{v}=n/4$
for trimerized defects.

It is evidently from Fig. \ref{Fig_conductivity} that different configurations
of N-related defects in graphene affect its conductivity both quantitatively
and qualitatively. Specifically, comparing conductivity for single
and double N substitutions in Figs. \ref{Fig_conductivity}(a) and
\ref{Fig_conductivity}(c) the following is noteworthy. For double
substitutions, the Fermi level is more strongly shifted away from
the Dirac point to the side of positive energies $E$ corresponding
to the $n$-type charge carriers. The dependence $\sigma=\sigma(n_{e})$
being sublinear for singly substituting N dopants transforms into
linear one for double substitutions. Double substitutions result to
the degradation of conductivity: particularly, $\sigma$ is decreased
by up to $\cong50\%$ for electron densities $n_{e}$ away from the
Dirac point (Fig. \ref{Fig_conductivity}(b)). However, conductivity
is not practically dependent on how pairs of N atoms are positioned:
as first-, second-, or third-nearest neighbours. (Note that abundant
amount of double substitutions of the second-nearest neighbouring
N atoms within the same sublattice have been confirmed experimentally
and theoretically in Ref. \citep{Lv2012}.)

Behaviours of conductivity in graphene with nitrogen--vacancy complexes
(Figs. \ref{Fig_conductivity}(b) and \ref{Fig_conductivity}(d))
are more complicated as compared with those seen in Figs. \ref{Fig_conductivity}(a)
and \ref{Fig_conductivity}(c) for graphitic defects, since the $\sigma=\sigma(n_{e})$
curves are dependent not only on the type of defects, their density,
and configuration, but on the type ($p$ or $n$) of charge carriers.
For electron densities $n_{e}$, all dependencies $\sigma=\sigma(n_{e})$
in Figs. \ref{Fig_conductivity}(b) and \ref{Fig_conductivity}(d)
are almost sublinear or superlinear for smaller (1\%) and larger (3\%)
defect concentrations, respectively; while for hole densities $-n_{e}>0$,
$\sigma=\sigma(n_{e})$ is linear, sublinear, or superlinear, depending
on the type of pyridine-like configurations. Such a behaviour is caused
by the more complex defect configurations due to the presence of simple
vacancies or divacancies. One can see from Fig. \ref{Fig_conductivity}(d)
that for hole densities ($-n_{e}>0$), which contribute to the dominant
$p$-type conductivity for pyridinic defects, the larger relative
content of vacancies ($n_{v}$), the degraded the conductivity for
the same total concentration of point defects ($n_{d}$). The same
effect is seen in Fig. \ref{Fig_conductivity}(b) at least for large
hole densities. These results agree with formation-energy (stability)
calculations \citep{Fujimoto2011} reported that graphite-like configuration
is the most stable; trimerized and tetramerized defects are less stable
and have small formation-energy difference; dimerized defects are
much less stable; at last, monomeric defects have the lowest stability
energy among the all N-doped defects in graphene.

The observed asymmetry of the conductivity curves with respect to
the neutrality point (Fig. \ref{Fig_conductivity}) follows from the
nature of the scattering potential, which impacts differently electrons
and holes due to its asymmetric property: $V<0$. The presence of
vacancies suppresses the electron--hole asymmetry in the conductivity.
This effect manifests itself for monomeric cases in Figs. \ref{Fig_conductivity}(b)
and \ref{Fig_conductivity}(d), where the relative vacancy concentration
with respect to the nitrogen one is the largest among the variety
of complex defects in Figs. \ref{Fig_Configurations}(e)--\ref{Fig_Configurations}(h).

\begin{figure*}
\includegraphics[width=1\textwidth]{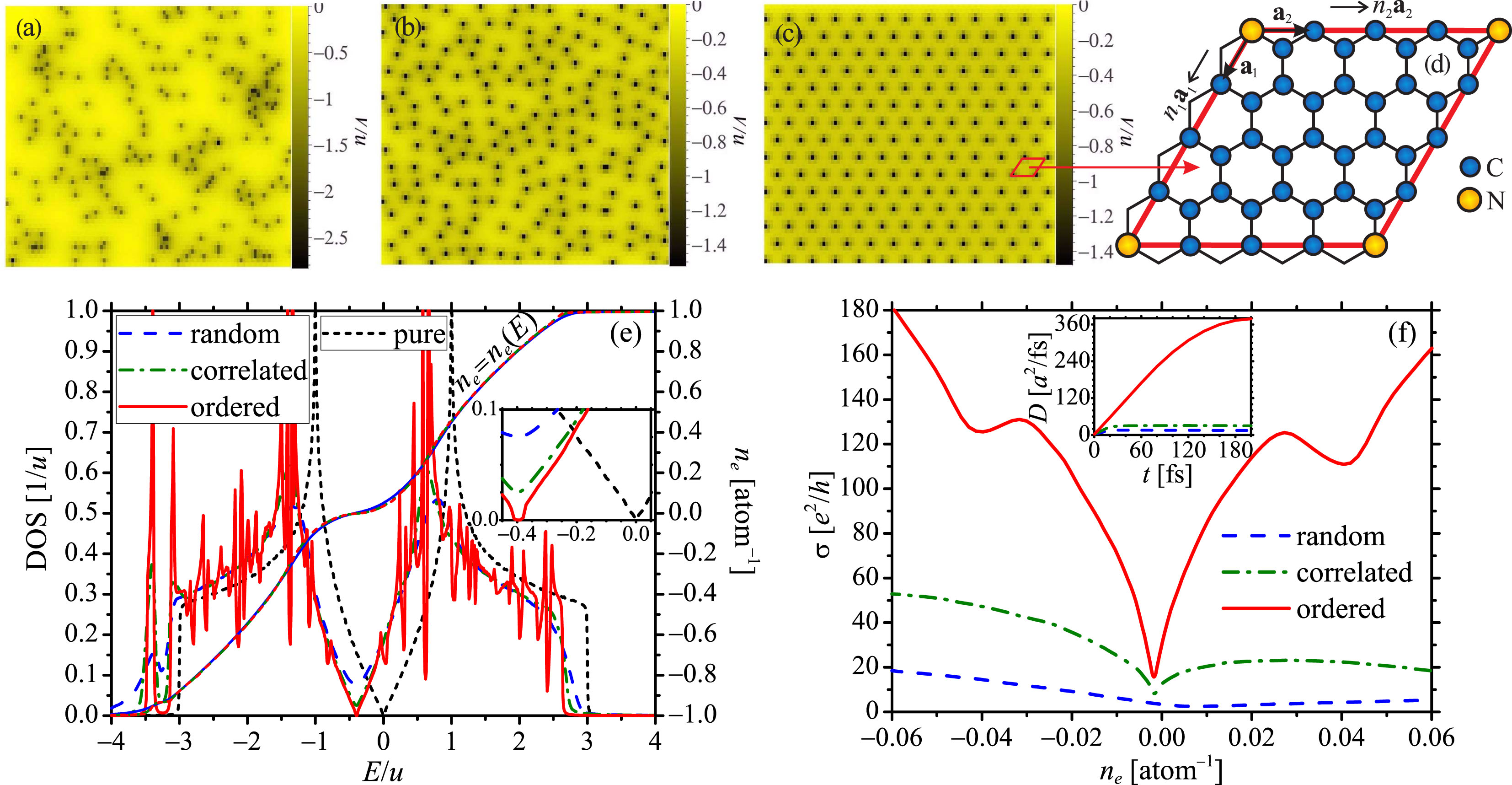}

\caption{(Colour online) (a)--(c) Scattering potential distributions, (e) density
of states (DOS), and (f) electron-density dependent conductivity for
$n_{\mathrm{N}}=3.125\%$ of (a) random, (b) correlated, and (c)--(d)
ordered nitrogen substitutions in graphene lattice. Ordered N atoms
(c) form superstructure with one N atom within the primitive unit
cell (d). Inset in (e): zoom of DOS in the band gap and zero energy
regions. Inset in (f): diffusivity vs. the time at the energy $E=0.2u$
for random, correlated, and ordered N atoms.}

\label{Fig_rnd-cor-ord}
\end{figure*}

Among the represented configurations of N-related defects in Fig.
\ref{Fig_Configurations}, substitutional (graphite-like) ones are
found to be the most common ($\cong75$--$80\%$ dominance among all
defects identified) \citep{Joucken2012,Lv2012}. Further, we deal
with a case of singly substituting N atoms for their random (Fig.
\ref{Fig_rnd-cor-ord}(a)), correlated (Fig. \ref{Fig_rnd-cor-ord}(b)),
and ordered (Fig. \ref{Fig_rnd-cor-ord}(c)) distributions over the
honeycomb-lattice sites. 

In case of correlation (Fig. \ref{Fig_rnd-cor-ord}(b)), N impurity
atoms are no longer considered to be randomly located. To describe
their spatial correlation, we adopt a model \citep{Li2011} using
the pair distribution function $p(\mathbf{R}_{i}-\mathrm{\mathbf{R}}_{j})\equiv p(r)$:

\begin{equation}
p(r)=\left\{ \begin{array}{l}
0,\; r<r_{0}\\
1,\; r\geq r_{0}
\end{array}\right.\label{p_corr}
\end{equation}
where $r=|\mathbf{R}_{i}-\mathbf{R}_{j}|$ is a distance between the
two N atoms, and the correlation length $r_{0}$ defines minimal distance
that can separate any two of them. Note, that for the randomly distributed
impurities, $r_{0}=0$. The largest distance $r_{0_{\mathrm{max}}}$
depends on the relative nitrogen concentration $n_{\mathrm{N}}$;
in our calculations for $n_{\mathrm{N}}=3.125\%$, we chose $r_{0}=r_{0_{\mathrm{max}}}=5a$. 

In case of ordering (Fig. \ref{Fig_rnd-cor-ord}(c)), we consider
structure in Fig. \ref{Fig_rnd-cor-ord}(d), where the relative content
of ordered N atoms is $n_{\mathrm{N}}=1/32=3.125\%$. This structure
forms a $\mathrm{C}_{31}\mathrm{N}$-type superstructure, where distribution
of N atoms over the honeycomb-lattice sites can be described by the
single-site occupation-probability function derived by the static
concentration wave method \citep{Radchenko-PhysE,Khachaturyan}. In
the computer implementation, $n_{\mathrm{N}}=3.125\%$ of N atoms
occupy sites within the same sublattice and can be described via a
single-site function:

\begin{equation}
P(\mathbf{R})=\left\{ \begin{array}{l}
1,\; n_{1}+n_{2}=4\mathbb{Z}\\
0,\;\mathrm{otherwise}
\end{array}\right.\label{P_ord}
\end{equation}
where $n_{1}$, $n_{2}$, and $\mathbb{Z}$ belong to the set of integers,
and $n_{1}$ and $n_{2}$ denote coordinates of sites in an oblique
coordinate system formed by the basis translation vectors $\mathbf{a}_{1}$
and $\mathbf{a}_{2}$ shown in Fig. \ref{Fig_rnd-cor-ord}(d).

Figure \ref{Fig_rnd-cor-ord}(e) shows the DOS and the electron density
$n_{e}=n_{e}(E)$ in graphene with $n_{\mathrm{N}}=3.125\%$ of random,
correlated, and ordered distributions of N impurity atoms. The Fermi
level is shifted with respect to $E=0$, which is related to negativity
of the scattering potential. The calculated DOS-curves for random,
correlated, and ordered N atoms are similar with different that peaks
appearing at $-4\lesssim E/u\lesssim-3$ for correlation and manifesting
themselves in whole energy interval for ordering. Such peaks in DOS
are due to the periodicity of scattering-potential distribution, which
describes ordered positions of N atoms in graphene-based superstructure
in Fig. \ref{Fig_rnd-cor-ord}(d). (Additional calculations \citep{Radchenko-NOVA}
showed that the peaks become stronger and even transform into discrete
energy levels with broadening as impurity concentration and/or periodic
potential increase.) Small band gap for ordered N configurations (see
inset in Fig. \ref{Fig_rnd-cor-ord}(e)) is caused by periodic potential
(Fig. \ref{Fig_rnd-cor-ord}(c)) resulting to ordered distribution
of N atoms over the sites of the same sublattice (Fig. \ref{Fig_rnd-cor-ord}(d)),
\textit{i.e.} to breaking of symmetry of two graphene sublattices.
Such a band gap opening due to superlattice of dopants has already
been discussed, \textit{e.g.}, in Refs. \foreignlanguage{english}{\citep{Park2008,Martinazzo2010,Casolo2011}}.
Belonging of ordered N dopants to one sublattice also leads to the
band gap appearance in graphene electronic spectrum even for a random
distribution of the N dopants \foreignlanguage{english}{\citep{Lherbier2013}}. 

In contrast to the cases of random and correlated N atoms, when steady
diffusive regime is reached, for the case of ordering a quasi-ballistic
regime is observed for a long time as it is shown in the inset of
Fig. \ref{Fig_rnd-cor-ord}(f). This (quasi-balistic) behaviour of
$D(t)$ indicates a very low scattered electronic transport, since
electrons live mainly in the sublattice that does not contain the
ordered N atoms \foreignlanguage{english}{\citep{Lherbier2013}}.
If the diffusive regime is not completely reached, the semiclassical
conductivity cannot be in principle defined. However we extracted
$\sigma$ for the case of ordered N atoms using the highest $D(t)$
when quasi-ballistic behaviour turns to a quasi-diffusive regime with
an almost saturated diffusivity coefficient.

One can see from Fig. \ref{Fig_rnd-cor-ord}(f) that correlation and
ordering of N dopants rise the conductivity up to several ($\cong3$--$6$)
and tens ($\cong20$--$30$) times, respectively, as compared with
their random distribution. We predict that such increasing of the
conductivity due to the spatial correlation of impurities should manifest
in a varying degree for any asymmetric ($V>0$ or $V<0$) scattering
potential modelling charged impurities or neutral adatoms. However,
for symmetric potential ($V\gtrless0$), correlation does not affect
the conductivity in graphene sheets \citep{Radchenko 1}. Note that
correlation effect (for potassium impurity atoms in graphene) have
been already observed in experiment \citep{YanFuhrer} and sustained
in theoretical predictions based on the standard semiclassical Boltzmann
approach within the Born approximation \citep{DasSarma2011}. Figure
\ref{Fig_rnd-cor-ord}(f) demonstrates that ordering of N impurity
atoms also suppresses the electron--hole asymmetry of the conductivity,
at least, for small charge-carrier densities.

\section{Conclusions}

The variety of different experimentally-revealed configurations of
nitrogen impurity atoms in both graphite- and pyridine-type defects
in graphene affects its conductivity quantitatively and qualitatively
that we numerically demonstrated using the quantum-mechanical Kubo--Greenwood
approach. Charge-carrier-density dependence of the conductivity, $\sigma=\sigma(n)$,
can be linear, sublinear, or superlinear depending on relative concentrations
of N dopants and vacancies, their configurations over the graphene-lattice
sites, and type of carriers---electrons or holes. For the most common
graphitic defects, N atoms doubly substituting C ones in graphene
degrade its conductivity as compared with their single substitutions.
The presence of vacancies in the complex defects as well as ordering
of substitutional N atoms suppresses the electron--hole asymmetry
of the conductivity. Correlated and ordered N scatterers in graphene
modelled by the negative scattering potential enhance the conductivity
up to the several and tens times, respectively, as compared with their
random distribution. Demonstrated influences of the various N-doping
configurations, especially ordering effect and herewith band gap opening,
in the N-doped graphene suggest the possibility of tailoring graphene
transport properties via the positioning neutral adatoms or charged
impurities on the substrate.

\section*{Acknowledgements}

T.M.R. benefited immensely from collaboration with Igor Zozoulenko
and Artsem Shylau, and thank Sergei Sharapov, Valery Gusynin, Vadim
Loktev, and Maksym Strikha for discussions. Yu.I.P. is grateful to
DAAD for support.


\begin{thebibliography}{10}
\bibitem{D_t} $D(E,t)$ has a dimension of diffusivity, nevertheless,
it is not a diffusion coefficient in a nondiffusive ballistic regime,
when there are no any scatterings.
\end{thebibliography}
\end{document}